\documentclass[preprint,review,12pt]{elsarticle}

\usepackage{amsmath}
\usepackage{tabularx}
\usepackage{longtable}
\usepackage{lineno,hyperref}
\usepackage{breqn}
\usepackage{amssymb}
\usepackage{graphicx}
\usepackage{multirow}
\usepackage{array}
\usepackage{amsmath,amssymb,amsfonts}
\usepackage{algorithmic}
\usepackage{algorithm}
\usepackage[table,xcdraw]{xcolor}
\usepackage{amsthm}
\usepackage{color}
\usepackage{soul}
\usepackage{ulem}
\usepackage{amsmath}
\usepackage{subcaption}
\usepackage{tabularx,xurl}
\usepackage{rotating}
\usepackage[margin=3cm]{geometry}

\modulolinenumbers[5]

\journal{Computer Communications}

\makeatletter
\g@addto@macro\normalsize{%
  \setlength\abovedisplayskip{2pt}
  \setlength\belowdisplayskip{2pt}
  \setlength\abovedisplayshortskip{2pt}
  \setlength\belowdisplayshortskip{2pt}
}
\makeatother

\begin{document}

\begin{frontmatter}

\title{End-to-End Delay Guaranteed SFC Deployment: \\ A Multi-level Mapping Approach}

\author{Fatemeh Yaghoubpour}
\ead{f.yaghoubpour@aut.ac.ir}
\author{Bahador Bakhshi\corref{cor1}}
\ead{bbakhshi@aut.ac.ir}
\author{Fateme Seifi}
\ead{fateme.seifi@aut.ac.ir}

\cortext[cor1]{Corresponding author}
\address{Amirkabir University of Technology, Hafez Avenue, Tehran, Iran}

\begin{abstract}
Network Function Virtualization (NFV) enables service providers to maximize the business profit via resource-efficient QoS provisioning for customer requested Service Function Chains (SFCs). In recent applications, end-to-end delay is one of the crucial QoS requirement needs to be guaranteed in SFC deployment. While it is considered in the literature, accurate and comprehensive modeling of the delay in the problem of maximizing the service provider’s profit has not yet been addressed. In this paper, at first, the problem is formulated as a mixed integer convex programming model. Then, by decomposing the model, a two-level mapping algorithm is developed, that at the first level, maps the functions of the SFCs  to virtual network function instances; and in the second level, deploys the instances in the physical network. Simulation results show that the proposed approach achieves a near optimal solution in comparison the performance bound obtained by the optimization model, and also superiors the existing solutions.
\end{abstract}


\end{frontmatter}

\section{Introduction}\label{Intro}
Recently, due to the advent of the Internet of Things (IoT) and 5G, Network Function Virtualization (NFV) has received much attention to address heterogeneous requirements of services. Since NFV decouples Network Functions (NFs) from the physical substrate network, it creates this opportunity for Telecommunication Service Providers (TSPs) to effectively upgrade their network to guarantee a wide range of QoS metrics, such as high data rate and reliability as well as low end-to-end delay and packet loss. That obviously reduces the Capital Expenses (CAPEX) as well as the Operational Expenses (OPEX). Moreover, this decoupling makes a drastic increase in agility and time-to-market of network services. To exploit its potential, extensive researches have been done on NFV-based networks during the last years \cite{hawilo2014nfv,han2015network,herrera2016resource}.

Resource allocation to customer services is the main challenge in NFV. In this context, the services are described as Service Function Chains (SFCs) containing a sequence of Virtual Network Functions (VNFs) connected with virtual links. To serve the customers, the TSP needs to deploy the SFCs by allocating the required resources, like processing, storage, and bandwidth. Service providers attempt to allocate the shared substrate network resources to SFCs efficiently to achieve objectives such as maximum gain, acceptance rate, reliability, and energy efficiency. This problem is named SFC mapping \cite{yu2017qos}.

Quality of Service (QoS) provisioning is the crucial requirement in SFC mapping, and end-to-end delay is one of its essential metrics which is considered in recent mission-critical and multimedia applications. End-to-end delay is a combination of heterogeneous delay factors, including propagation, transmission, processing, and queuing delays. Accordingly, not only modeling the different factors but also provisioning the requirement in such a complex network is a serious challenge in SFC mapping.

In recent years, the SFC mapping problem is concerned by network researchers, and several papers have investigated it from different aspects \cite{hawilo2014nfv, herrera2016resource, mijumbi2015network}. However, even though delay is one of the most significant requirements of SFCs, it has not been addressed significantly. In \cite{xie2020online,fischer2019construction}, and \cite{qu2017reliability} the end-to-end delay is calculated by accumulating the propagation delay and other factors such as transmission, and queuing delays are ignored. In \cite{tahmasebi2018optimum}, only processing delay is concerned and the other factors are neglected. Although authors in \cite{woldeyohannes2018cluspr} consider propagation and processing delays, they do not pay attention to queuing and transmission delay. In \cite{alameddine2017interplay}, the authors calculate processing and transmission delays; however, they ignore the queuing and propagation factors. In  \cite{bhamare2017optimal}, processing, transmission, and queuing delays are considered in a cloud environment, and in \cite{chen2018automated}, queue length is considered as a queuing delay metric.

The propagation, transmission, and processing delays can be formulated as linear functions, while the queuing delay depends on the traffic distribution, and in the simplest form, is a fractional function that makes the model non-linear. Hence, for the sake of simplicity, most researchers tend to eliminate it. However, queuing delay is one of the important factors in end-to-end delay and mostly is not negligible in comparison to the other factors. Therefore, considering queuing delay is inevitable. In this paper, for this purpose, we define a  convex model that takes into account propagation, transmission, processing, and queuing delays as the constraints with the aim of maximizing the business profit of the TSP. To the best of our knowledge, the problem of gain maximization subject to assuring the end-to-end delay considering all the affecting factors, which is named Delay Constrained SFC Mapping (DCSM), has not been addressed in the literature yet. Hereupon, for filling this research gap, we model the problem as a mixed-integer convex problem and develop an efficient algorithm to solve it. This paper extends our previous work \cite{yaghoubpour2019optimal}.

In the DCSM problem, there is a set of input SFCs that the service provider can whether or not accept them. In addition to processing and bandwidth resources, each SFC has its end-to-end bandwidth requirement. The objective is gain maximization while satisfying the end-to-end delay in terms of propagation and queuing delays. Briefly, our contributions comparing to the previous researches are:
\begin{itemize}
\item For the first time, we formulate the DCSM problem as a mixed integer convex model.
\item We develop a multi-level mapping algorithm by decomposing the optimization model.
\item We evaluate the proposed solution in different setting against to theoretical performance bound and existing solutions. 
\end{itemize}

The remainder of this paper is organized as follows. In Section \ref{RW}, we review related work, categorize them, and identify the research gap. In Section \ref{SMPF}, first, the system model is explained; afterward, DCSM is formulated as a mixed integer convex optimization problem. The optimization model is decomposed and a multi-level mapping algorithm is presented in Section \ref{PS} to solve the problem efficiently. Section \ref{SR} evaluates the simulation results. Finally, we will make a conclusion and introduce some new ideas as future work in Section \ref{CON}.

\section{Related Work}\label{RW}

In this section, related work to the DCSM problem are reviewed, the differences with this paper are investigated, and finally, the research gap is identified.

In \cite{woldeyohannes2018cluspr}, the authors investigated resource allocation scheme for  balancing the load among NF instances, and maximizing the total network utilization. They considered CPU, disk, and memory of physical nodes and bandwidth capacity of physical links. For end-to-end delay, only  propagation and processing delays were assumed; and the authors proposed a heuristic algorithm. This work ignores the  queuing delays and also does not consider the license costs of VNF instance and the cost of physical resources. 
In the SFC mapping problem in \cite{nejad2018vspace}, physical nodes and links capacity limitations are considered. However, VNF instances are not shared among  chains and no end-to-end delay is guaranteed. Moreover, the authors ignored the licensing and physical resources activating costs.

In \cite{xie2020online}, an ILP\footnote{Integer Linear Programming}  model is proposed to address the chain deployment problem considering the delay of the chain and the maximum load among edge devices. Only CPU is considered as the physical node resource while disk and memory are ignored. Moreover, only the propagation delay is considered. 
In \cite{alameddine2017interplay}, a formulation of the SFC scheduling (SFCS) problem is presented that exploits interactions between NFs mapping onto VNFs, service scheduling, and traffic routing. Physical nodes and virtual instances are shared among multiple SFCs. The authors assume that all SFCs are accepted without considering any cost for provisioning.
The authors in \cite{bhamare2017optimal}, proposed an ILP model that considers CPU, disk, and bandwidth constraints. The processing, transmission, and queuing delays are addressed in a cloud environment. However, VNF instances are not shared among chains and it is assumed that all requested chains are accepted, i.e., there is no CAC (Call Admission Control) mechanism. This paper proposes an affinity based heuristic algorithm to solve the problem.

The work presented in \cite{fischer2019construction}, which is the improved version of \cite{bhamare2017optimal}, assigns fixed delays to physical nodes and physical links. The work aims at cost minimization and considers physical nodes and links costs; however, it neglects the cost of activating VNF instances as well as the transmission and queuing delays. This paper does not have any CAC mechanism and assumes that all SFCs have to be accepted. A heuristic algorithm named Annealing Based Simulation Approach (ABSA) is proposed, wherein the affinity-based algorithm proposed in \cite{bhamare2017optimal} is used.
The SFC mapping problem to minimize the time-average cost is formulated as a MILP model in \cite{chen2018automated}. This paper considers CPU and bandwidth as the substrate network resource. It aims at minimizing the lengths of queues in order to decrease the end-to-end delay. Both physical nodes and virtual instances are shared among multiple SFCs.

The objective of the problem proposed in \cite{tahmasebi2018optimum} is delay minimization, but the transmission delay is the only considered metric. It only considers physical links bandwidth as the constraint. It shares the physical nodes among multiple chains. The CAC problem and cost are also considered.
In \cite{qu2017reliability}, the objective is solving the SFC mapping problem such that satisfies the reliability and end-to-end delay requirements while minimizing the communication bandwidth usage. It just considers CPU and bandwidth resources, and does not share instances among multiple chains. The only considered delay metric is the propagation delay.
In \cite{yuan2020delay}, a system model based on deep reinforcement learning (DRL) is defined to minimize the overall network delay while ensuring fairness among different flows. It calculates queuing delays and assumes it is the only parameter affecting the end-to-end delay. It does not pay attention neither to CAC nor VNF instance sharing.

In paper \cite{gouareb2018virtual}, the problem of VNF placement is studied with the objective of minimizing the overall delay. To do so, it assigns a list of shortest path to each chain using Dijkstra’s algorithm \cite{makariye2017towards} in a weighted network. It considers queuing and propagation delays. It considers CPU, memory, and bandwidth as physical resource restrictions. It shares physical nodes and virtual instances among different SFCs. In this paper, no CAC mechanism is considered. Paper \cite{kar2017energy} formulates an energy cost optimization problem considering resource capacities restriction and satisfying a delay threshold as constraints. Queuing delay is the only considered delay metric. This problem assumes that all SFCs are accepted and there is no CAC problem. Energy consumption is the cost metric.

Table \ref{tt1} summarizes the discussed works related to the DCSM problem. The columns of the table identify the categorizing criteria and the last row is this paper. Comparing the previous work, we can conclude that the problem of gain maximization while modeling the end-to-end delay regarding queuing, processing, transmission, and propagation delays, and sharing the instances among multiple SFCs is a research gap that we will go through it in this paper. 

\begin{table} 
\begin{center}
	\caption{Summary of Related Work}
	\label{tt1}
	\centering
	\small\addtolength{\tabcolsep}{0pt}
	\scalebox{0.9}{
	\begin{tabular}{|c|c|c|c|c|c|c|c|c|c|c|c|c|c|c|}

  \hline
  \multirow{2}{*}{{Paper}} & \multicolumn{4}{|c|}{Delay}& \multicolumn{5}{|c|}{Capacity Constraint}&\multicolumn{2}{|c|}{Sharing}& \multirow{2}{*}{\rotatebox{90}{CAC}}&\multirow{2}{*} {\rotatebox{90}{{Optimization Model}}}& \multirow{2}{*}{\rotatebox{90}{Heuristic Algorithm}}\\ \cline{2-12}
  &\rotatebox{90}{Propagation}& \rotatebox{90}{Transmission}& \rotatebox{90}{Processing}& \rotatebox{90}{Queuing}&\rotatebox{90}{CPU}& \rotatebox{90}{Storage}& \rotatebox{90}{Memory}& \rotatebox{90}{Networking}& \rotatebox{90}{Bandwidth}&\rotatebox{90}{Physical Nodes}& \rotatebox{90}{VNF Instance}&&&\\
  \hline
  \hline

	\cite{woldeyohannes2018cluspr} & $\surd$ & $\surd$ & $\surd$ & {\ } & $\surd$ & $\surd$ & $\surd$ & {\ } & $\surd$ & $\surd$ & $\surd$ & $\surd$ & $\surd$ & $\surd$
	\\
	\hline

	\cite{nejad2018vspace} & {\ } & {\ }& {\ }& {\ }& $\surd$& $\surd$& $\surd$& {\ }& $\surd$& $\surd$& {\ }& $\surd$& $\surd$& $\surd$
	\\
	\hline

	\cite{xie2020online} & $\surd$ & {\ }& {\ }& {\ }& $\surd$& {\ }& {\ }& {\ }& $	\surd$& $\surd$& $\surd$& $\surd$& $\surd$& $\surd$\\
	\hline

	\cite{alameddine2017interplay} & {\ } & $\surd$& $\surd$& {\ }& $\surd$& {\ }& {\ }& {\ }& $\surd$& $\surd$& $\surd$& {\ }& $\surd$& $\surd$ \\
	\hline

	\cite{bhamare2017optimal} & {\ } & $\surd$& $\surd$& $\surd$& $\surd$& {\ }& {\ }& $\surd$& $\surd$& $\surd$& {\ }& {\ }& $\surd$& $\surd$\\
	\hline

	\cite{fischer2019construction} & $\surd$ & {\ }& $\surd$& {\ }& $\surd$& $\surd$& {\ }& $\surd$& $\surd$& $\surd$& {\ }& {\ }& $\surd$& $\surd$\\
	\hline

	\cite{chen2018automated} & {\ } & {\ }& {\ }& $\surd$& $\surd$& $\surd$& {\ }& {\ }& $\surd$& $\surd$& $\surd$& {\ }& $\surd$& $\surd$\\
	\hline

	\cite{tahmasebi2018optimum} & {\ } & {\ }& $\surd$& {\ }& {\ }& {\ }& {\ }& {\ }& $\surd$& $\surd$& {\ }& $\surd$& {\ }& {\ }\\
	\hline

	\cite{qu2017reliability} & {\ } & {\ }& {\ }& {\ }& $\surd$ & {\ }& {\ }& {\ }& $\surd$& $\surd$& {\ }& $\surd$& $\surd$& $\surd$\\
	\hline

	\cite{yuan2020delay} & {\ } & {\ }& {\ }& $\surd$& $\surd$ & {\ }& {\ }& {\ }& $\surd$& $\surd$& {\ }& {\ }& $\surd$& $\surd$ \\
	\hline

	\cite{gouareb2018virtual} & $\surd$ & {\ }& {\ }& $\surd$& $\surd$ & {\ }& $\surd$& {\ }& $\surd$& $\surd$& $\surd$& {\ }& $\surd$& $\surd$ \\
	\hline

	\cite{kar2017energy} & {\ } & {\ }& {\ }& $\surd$& $\surd$ & {\ }& {\ }& {\ }& $\surd$& $\surd$& $\surd$& {\ }& $\surd$& $\surd$ \\
	\hline

	\cite{yaghoubpour2019optimal} & $\surd$ & {\ }& $\surd$& $\surd$& $\surd$ & $\surd$& $\surd$& $\surd$& $\surd$& $\surd$& $\surd$& $\surd$& $\surd$& {\ }\\
	\hline

	This paper & $\surd$ & {\ }& $\surd$& $\surd$& $\surd$ & $\surd$& $\surd$& $\surd$& $\surd$& $\surd$& $\surd$& $\surd$ & $\surd$ & $\surd$\\
	\hline

	\end{tabular}	
	}		
\end{center}
\end{table}



\section{System Model AND Problem Statement} \label{SMPF}
In this section, we first discuss the assumptions and present the system model, then we state the problem with an illustrative example, and finally formulate it.

\subsection{Assumptions and System Model}
In this paper, $G_p= \left(N_p,E_p\right)$ demonstrates the substrate network wherein $N_p$ and $E_p$ respectively represent the physical nodes and links. Each $n\in N_p$ indicates a physical server wherein $\theta_n^{CPU}$,$\theta_n^{mem}$, and $\theta_n^{strg}$ denote its total available CPU, memory, and storage, respectively. Additionally, link $\left(n,m\right)\in E_p$ shows a physical connection between two physical nodes $n,m \in N_p$. This link has a limited data transfer capacity $\theta_{\left(n,m\right)}$ and propagation delay $d_{\left(n,m\right)}$.

Set $T$ contains all available VNF types that can be requested by users. In case of requesting a type in a chain, the service provider activates an instance of the VNF type by running the image of the VNF on a virtual machine and installing the required licenses. $I_t$ represents all available instances of type $t\in T$ that can be activated on-demand. The instances of each type have three specific characteristics. The first one is the required physical resources (CPU, storage, and memory) that need to be allocated for activating each instance of the type. The second characteristic is the traffic processing capacity of the instances of that type. These two characteristics are determined by the VNFs vendors wherein the CPU, memory, and storage requirements of each instance of type $t\in T$ are denoted by ${CPU}_t$, ${mem}_t$, and ${strg}_t$ respectively, and the traffic processing capacity of the type is indicated by $\mu_t$. At last, the third characteristic of type $t\in T$  is the license fee, i.e., the amount of money that the service provider should pay to the VNF vendor to activate a license of that type; it is denoted by $C_t$. It worth mentioning that we assume that each instance has the sufficient capacity to serve at least one chain, and therefore a VNF of a chain is never split among multiple instances. Similarly, we assume that each instance also has to be mapped exactly one one individual physical server, and it cannot be split among multiple servers.

All SFCs requested by users are gathered in set $G_v$. Each requested SFC $r \in G_v$, is defined as a sequence of VNFs which are connected by virtual links. Set $N_v^r$ contains all VNFs of SFC $r$ and $E_v^r$ contains all directed virtual links connecting the sequence of the VNFs of the SFC. Parameter $w_{\left(u,v\right)}^r$ denotes the required bandwidth for virtual link $\left(u,v\right)\in E_v^r$ connecting two consecutive VNFs $u,v \in N_v^r$ of SFC $r$. As mentioned, each VNF $u \in N_v^r$ has a specific type which is denoted by  $t_u^r \in T$.  Table \ref{tt2} summarizes the notations used throughout this paper.

\begin{table} 
\begin{center}
\caption{Notations and Definitions}
\label{tt2}
\centering
\small\addtolength{\tabcolsep}{0pt}
\scalebox{0.9}{
\begin{tabular}{|c|l|}
 
  \hline
  Notation & Definition \\
  \hline
  \hline
  $ G_{p}$ & Infrastructure network graph\\
  \hline
  $N_{p}$ & The set of substrate nodes\\
  \hline
  $E_{p}$ & The set of substrate links\\
  \hline
  ${\theta _{\left( {n,m} \right)}}$ & The available capacity of physical link $\left( {n,m} \right)\in {E_p}$\\
  \hline
  $\theta_{n}^{CPU}$ & The available CPU cores of physical node $n \in {N_p}$ \\
  \hline
  $\theta_{n}^{mem}$ & The available memory of physical node $n \in {N_p}$ \\
  \hline
  $\theta_{n}^{strg}$ & The available storage of physical node $n \in {N_p}$ \\
  \hline
  ${\mu _n}$ & Processing rate of physical node $n \in {N_p}$\\
  \hline
   $ d_{\left( {n,m} \right)}$ & The propagation delay of link $\left( {n,m} \right) \in E_p $ \\
  \hline
    ${\rho}$ & Bandwidth fee of physical link\\
  \hline
    ${\sigma}_n$ & The cost paid for activating physical node  $n \in N_p$\\
  \hline
  $G_{v}$ & Set of all requested service function chains\\
  \hline
  $N_{v}$ & Set of all VNFs of all SFCs\\
  \hline
  $E_{v}$ & Set of all virtual links of all SFCs\\
  \hline
  $N^{v}_{r}$& Set of all VNFs of SFC $r \in {G_v}$ \\
  \hline
  $E^{v}_{r}$& Set of all virtual links of SFC  $r \in {G_v}$ \\
  \hline
  $d_{r}^{th}$ & The maximum end-to-end delay threshold of SFC $r \in G_v$\\
  \hline
  $f_{r}$ & The input traffic of each VNF $u \in N^r_v$\\
  \hline
  $rev_{r}$ & The revenue of SFC  $r \in G_v$\\
  \hline
  $w_{\left( {k,l} \right)}^{r}$ & The required bandwidth of link $\left( {k,l} \right) \in E^r_v$  \\
  \hline
  $T$ & Set of all available VNF types  \\
  \hline
  $I_t$ & Set of available instances of types $t \in T$  \\
  \hline
  $\mu_t$ & Traffic capacity of type $t \in T$  \\
  \hline
  $C_t$ & The cost that the provider pays for the activation of each instance of type $t \in T$  \\
  \hline
  $CPU_t$ & The required CPU of each instance of type $t \in T$  \\
  \hline
  $mem_t$ & The required memory of each instance of $t \in T$  \\
  \hline
  $strg_t$ & The required storage of each instance of $t \in T$  \\
  \hline
  $t_u^r$ & The type of VNF $u \in N_v^r$ of chain $r \in G_v$  \\
  \hline
\end{tabular}	
}		
\end{center}
\end{table}

In this paper, similar to \cite{bhamare2017optimal, chen2018automated, gouareb2018virtual, yuan2020delay, kar2017energy, han2019utility, erel2018take, prados2019performance}, we consider the \textit{average} end-to-end delay between the source and the destination of each service chain as the QoS metric. Threshold $d_r^{th}$ indicates the utmost average tolerable end-to-end delay of SFC $r$. 

The following assumptions are made about the business model of the service provider. The provider’s revenue is made by accepting SFCs which is $R_r$ for SFC $r$. On the other hand, it should pay license cost for activating the VNF instances as well as the cost of allocating bandwidth on physical links as it is assumed that the physical links are leased from another transmission carrier network.
Here, we assume that to minimize the cost of service provider, the activated VNF instances in the network can be shared among multiple chains subject to the traffic processing capacity of the instance.

In this paper, we consider propagation, processing, and queuing delays as the most significant parameters imposing the end-to-end delay. 
In each physical node, the processing and queuing delays are induced by the hypervisor of the machine and also the VNF instances running on the hypervisor. More precisely, each packet at each node:

\begin{enumerate}
\item enters the physical node’s hypervisor queue,
\item is processed by the hypervisor and conducted to the virtual machine running the corresponding VNF,
\item enters the queue of the virtual machine, and
\item takes the required service from the VNF instance.
\end{enumerate}

\begin{figure}
  \centering
  \captionsetup{justification=centering}
  \includegraphics[width=0.4\textwidth]{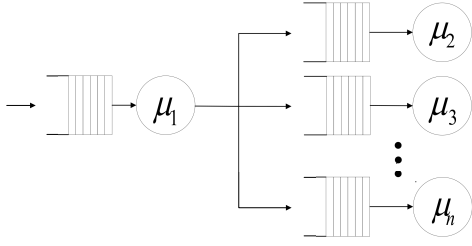}
  \caption{Queuing network at each physical node.}\label{Fig1}
\end{figure}

We model the delay induced by these steps as a queuing network depicted in figure \ref{Fig1}, where the left-hand server is the hypervisor and the right-hand servers are the VNF instances. The queues in this network are assumed as an $M/M/1$ system \cite{bolch2006queueing}, wherein the arrival process is Poisson, the service time is distributed exponentially, and there is just a single server processing the incoming traffic. Therefore, the average delay in each queue is
\begin{equation}\label{c1}
\bar d = \frac{1}{{\mu  - \lambda }}
\end{equation}
where, $\lambda$ and $\mu$ are respectively, the arrival rate, the departure rate of the packets.

For physical node $n \in N_p$, the arrival rate into the hypervisor queue, shown by $\lambda_n$, equals to the summation of all incoming flows to the node, and the service rate of the hypervisor, shown by $\mu_n$, is a given parameter. Therefore, the average delay of the hypervisor is:
\begin{equation}\label{c2}
{\bar d_n} = \frac{1}{{{\mu _n} - {\lambda _n}}}.
\end{equation}
By taking logarithm from \eqref{c2}, we have
\begin{equation}\label{c3}
\log \left( {{{\bar d}_n}} \right) + \log \left( {{\mu _n} - {\lambda _n}} \right) = 0,
\end{equation}
which is more convenient to hand it in the optimization formulations.

In a similar way, the average delay in each instance $i$ of type $t$, denoted by $\bar d_{i,t}$, is 
\begin{equation}\label{c4}
{\bar d_{i,t}} = \frac{1}{{{\mu _t} - {\lambda _{i,t}}}},
\end{equation}
where $\lambda _{i,t}$ and $\mu_t$ are  the total flow rate and the processing rate of of the instance respectively. And, again by taking logarithm from \eqref{c4}, we have
\begin{equation}\label{c5}
\log \left( {{{\bar d}_{i,t}}} \right) + \log \left( {{\mu _t} - {\lambda _{i,t}}} \right) = 0.
\end{equation}

\subsection{Problem Statement}

\begin{figure}
  \centering
  \captionsetup{justification=centering}
  \includegraphics[width=0.6\textwidth]{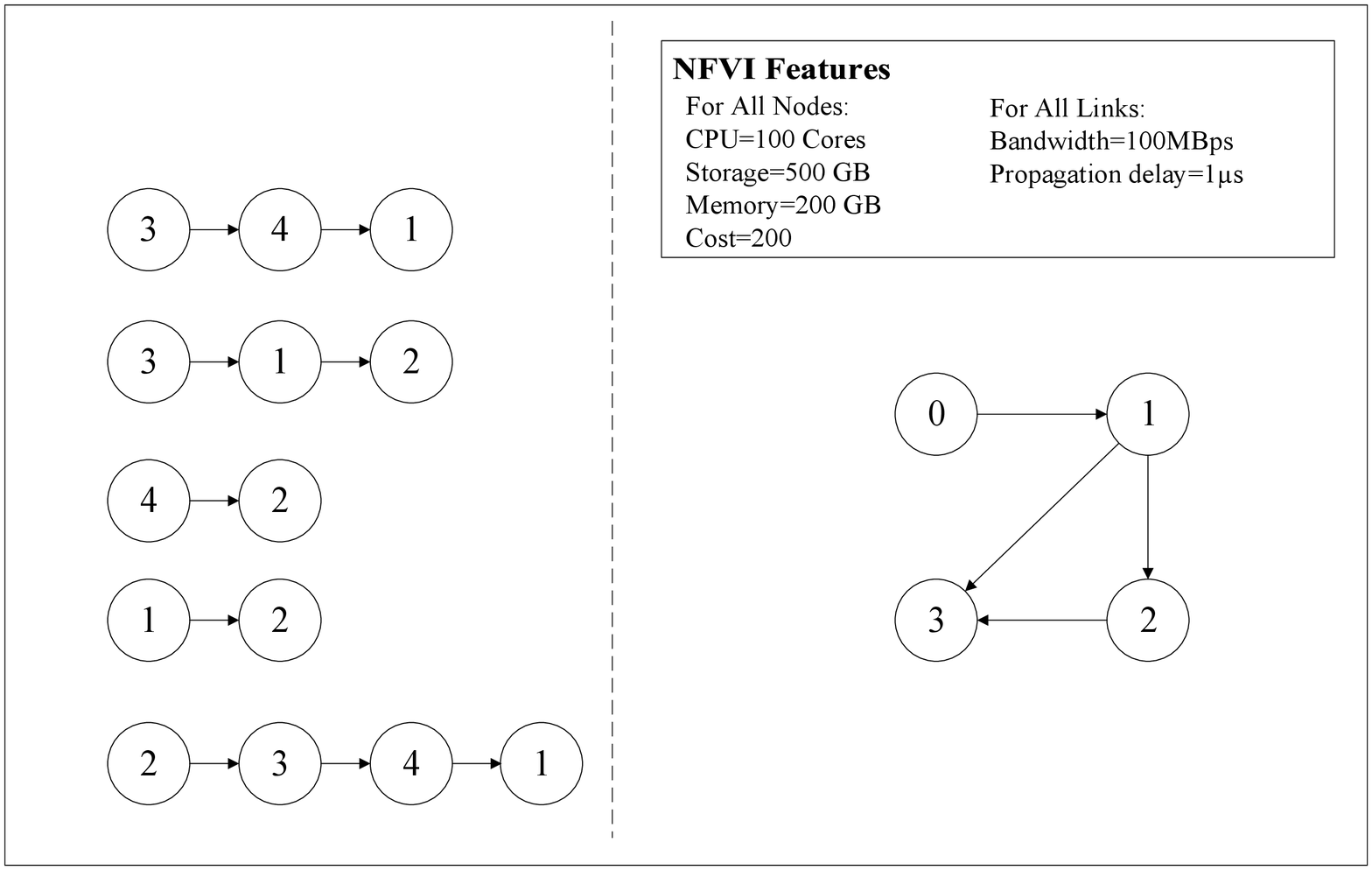}
  \caption{The inputs of the  DCSM problem. The left side is the set of SFCs with a specified type per VNF. The right side is the substrate network.}\label{Fig2}
\end{figure}

In this paper, we investigate the  DCSM problem, where, a service provider, who operates the infrastructure network $G_p$, aims to deploy the set $G_v$ of SFCs in order to maximize its profit while satisfying the SFCs requirements. The profit is a function of the revenue of accepting the demands as well as the cost paid for VNFs license fee. The SFCs requirements include not only processing and bandwidth but also the end-to-end delay which is a function of processing, queuing and propagation delays. In the following, we clarify the problem by an illustrative example shown in Figure \ref{Fig2}. 

In this figure, the left side represents 5 input SFCs which are composed of different types of VNFs, indicated by the numbers inside the VNFs; and the right side represents the substrate network topology. Table \ref{tt3} and Table \ref{tt4} respectively show the settings of the SFCs and VNF types in this example. 

\begin{figure}
  \centering
  \captionsetup{justification=centering}
  \includegraphics[width=0.6\textwidth]{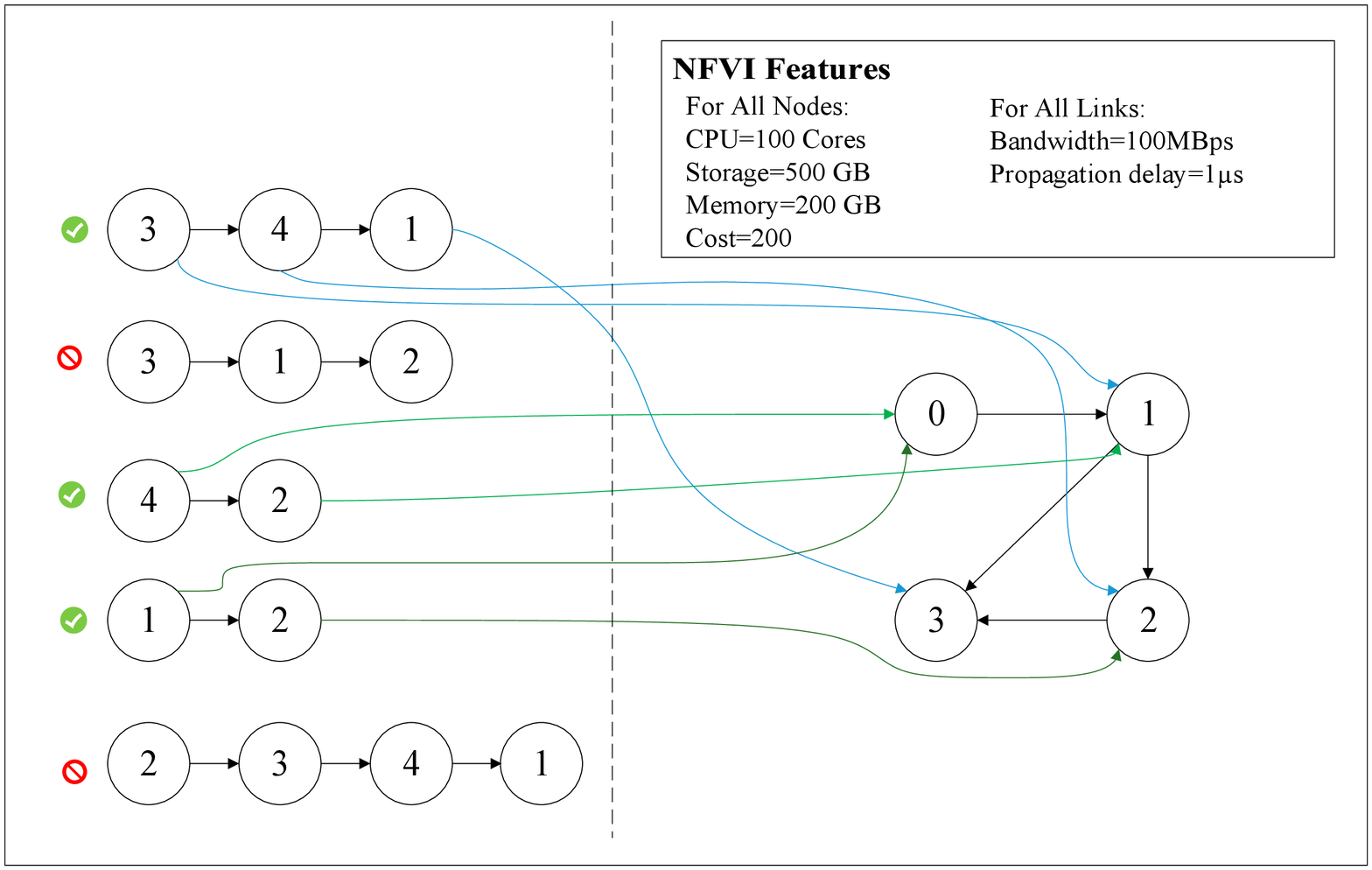}
  \caption{The accepted SFCs and their mappings.}\label{Fig3}
\end{figure}

The optimal solution of this instance of the DCSM problem, which is obtained by solving the optimization model proposed in the next section, is depicted in Figure \ref{Fig3} and Table \ref{tt5}. Figure \ref{Fig3} shows which SFCs are accepted and how they are mapped to the substrate network and Table \ref{tt5} illustrates the end-to-end delay that the accepted SFCs tolerate. As it shown, to maximize the objective function and to meet the SFC requirements, two SFCs are rejected. 
 

\begin{table} 
\begin{center}
\caption{The settings of the SFCs in Figure \ref{Fig2}}
\label{tt3}
\centering
\small\addtolength{\tabcolsep}{0pt}
\scalebox{0.9}{
\begin{tabular}{|c|c|c|c|c|c|}

  \hline
   {$r$} &{$|N^v_r|$}&{$t^r_u$}&{$f_r$}&{$rev_r$}&{$d_r^{th}$}\\
  \hline
  \hline
0&3&4,1,3&20&1050&0.3\\
\hline
1&3&1,2,3&30&1050&0.3\\
\hline
2&2&2,4&30&700&0.2\\
\hline
3&2&2,1&40&700&0.2\\
\hline
4&4&3,4,1,2&40&1400&0.4\\
\hline
\end{tabular}	
}		
\end{center}
\end{table}

\begin{table} 
\begin{center}
\caption{The settings of the VNF types in Figure \ref{Fig2}}
\label{tt4}
\centering
\small\addtolength{\tabcolsep}{0pt}
\scalebox{0.9}{
\begin{tabular}{|c|c|c|c|c|c|}
 
   \hline
   
   $t$ & $C_{t}$ & $CPU_{t}$ & $mem_{t}$ &  $strg_{t}$ & $\mu_{t}$ \\
  \hline
  \hline
  
1&100&100&150&300&100\\
\hline
2&150&100&150&300&100\\
\hline
3&120&100&150&300&100\\
\hline
4&140&100&150&300&100\\
\hline
\end{tabular}	
}		
\end{center}
\end{table}

\begin{table} 
\begin{center}
\caption{The components (hypervisor, VNF instances, and propagation) of the end-to-end delay of the accepted demands in the illustrated example of Figure \ref{Fig2}}
\label{tt5}
\centering
\small\addtolength{\tabcolsep}{0pt}
\scalebox{0.9}{
\begin{tabular}{|c|c|c|c|}
  \hline
   $r$ & $\sum{\bar{d}_n}$ & $\sum{\bar{d}_{i,t}}$ & $\sum{d_{(n,m)}}$\\
  \hline
  \hline
0&0.004547&0.053332&0.000001\\
\hline
2&0.004597&0.058333&0.000002\\
\hline
3&0.006578&0.057499&0.000002\\
\hline
\end{tabular}	
}		
\end{center}
\end{table}

\subsection{Problem Formulation}

In this section,  the DCSM problem is formulated. At the begging, we introduce the decision variables of the model, shown in table \ref{tt6}, and inter-dependencies between them; then we discuss the constrains and objective function of the model.

\begin{table} 
\begin{center}
\caption{Decision Variables and Corresponding Definitions}
\label{tt6}
\centering
\small\addtolength{\tabcolsep}{0pt}
\scalebox{0.9}{
\begin{tabular}{|p{0.15\linewidth}|p{0.85\linewidth}|}
  \hline
  Variable & Definition \\
  \hline
  \hline
  $ A_{r} \in {\{ {0,1} \}}$ & $A_r=1$ if and only if SFC $r \in G_v$ is accepted.\\
  \hline

  $ s_{i,t}^{u,r} \in {\{ {0,1} \}}$ & $s_{i,t}^{u,r}=1$ if and only if VNF $u \in N_r^v$ from SFC $r \in G_v$ is mapped to the instance $i \in I_t$ of type $t \in T$.\\
  \hline

  $ p_{n}^{i,t} \in {\{ {0,1} \}}$ & $p_{n}^{i,t}=1$ if and only if instance $i \in I_t$ of type $t \in T$ is mapped to the physical node $n \in N_p$.\\
  \hline

  $ z_{n,i}^{u,r} \in {\{ {0,1} \}}$ & $z_{n,i}^{u,r}=1$ if and only if VNF $u \in N_v^r$, is mapped to instance $i \in I_{t_u }$ running on physical node $n \in N_p$.\\
  \hline

  $ a_{n}^{r} \in {\{ {0,1} \}}$ & $a_{n}^{r}=1$ if and only if SFC $r \in G_v$ passes through physical node $n \in N_p$.\\
  \hline

  $ {\delta_{\left( n,m \right)}^{r,{\left( u,v \right)}}} \in R^{+} $ & $ {\delta_{\left( n,m \right)}^{r,{\left( u,v \right)}}}$ indicates the percentage of the traffic of the virtual link $ \left( u,v \right) \in E_v^r$ that is mapped to physical link $ {\left( n,m \right)} \in E_p$.\\
  \hline

  $ {y_n ^{i,t}} \in R^{+}$ & $y_n^{i,t}$ denotes the traffic of the hypervisor queue at node $n \in N_p$ which takes service from instance $i \in I_t$ of type $t\in T$.\\
  \hline

  $ {\bar{d_n}} \in R^{+}$ & $\bar{d_{n}}$ denotes the average queuing and processing delays of the hypervisor in node $n \in N_p$.\\
  \hline

  $ {\bar{d_{i,t}}} \in R^{+}$ & $\bar{d_{i,t}}$ denotes the average queuing and processing delays of instance $i\in I_t$ of type $t\in T$.\\
  \hline

  $ {d_{r}^{hyp}} \in R^{+}$ & $d_{r}^{hyp}$ indicates the imposed delay to SFC $r\in G_v$ by all the hypervisors the SFC passes through them.\\
  \hline

  $ {d_{r}^{ins}} \in R^{+}$ & $d_{r}^{ins}$ indicates the imposed delay to SFC $r \in G_v$ by all the instances servicing it.\\
  \hline

  $ {d_{r}^{link}} \in R^{+}$ & $d_{r}^{link}$ indicates the imposed delay to SFC $r\in G_v$ by all the  links forwarding its traffic.\\
  \hline

  $ {\lambda_{i,t}} \in R^{+}$ & $\lambda_{i,t}$ denotes the traffic entering instance $i\in I_t$ of type $t\in T$.\\
  \hline

  $ {l_{n}^{r}} \in R^{+}$ & $l_{n}^{r}$ indicates the delay that SFC $r\in G_v$ tolerates for passing through physical node $n\in N_p$.\\
  \hline

  $ {bw_{\left( n,m \right)}} \in R^{+}$ & $bw_{\left( n,m \right)}$ indicates the allocated bandwidth of physical link $\left(n,m\right) \in E_p$ to the accepted chains crossing the link.\\
  \hline

  $ {q_{i,t}^{u,r}} \in R^{+}$ & ${q_{i,t}^{u,r}}$ denotes the delay that SFC $r\in G_v$ tolerates since its VNF $u\in N_v^r$ is mapped to instance $i\in I_t$ of type $t\in T$.\\
  \hline

  $ {cost} \in R^{+}$ & $cost$ denotes the total money payed by the service provider for bandwidth allocation and instance activation.\\
  \hline

  $ {rev} \in R^{+}$ & $rev$ is the amount of money that the provider takes for the accepted SFCs\\
  \hline

  $ {gain} \in R$ & $gain$ is the difference between total revenue and cost and indicates the provider’s profit.\\
  \hline
\end{tabular}
}
\end{center}
\end{table}

\subsubsection{Decision Variables}
In this problem, there is a set of requested SFCs which the provider can either accept or not. Binary variable $A_r$ is for call admission control $\left( CAC \right)$ that denotes whether SFC $r$ is accepted, $A_r=1$. In this paper, by assuming the instance sharing capability, we propose a novel two-level-hierarchical mapping model for the DCSM problem. The first level is the mapping of the VNFs to a set of instances, and the second level is the mapping of the set of activated instances to the physical nodes. The binary variable $s_{i,t}^{u,r}$ denotes whether or not VNF $u \in N_v^r$ from SFC $r$ is mapped to the instance $i \in I_t$ of type $t \in T$. Similarly, $p_n^{i,t}$  denotes whether  the instance $i \in I_t$ of type $t \in T$ is mapped to the physical node $n \in N_p$.

It is necessary to indicate if a VNF is mapped to a physical node which is achieved in two steps. At first, binary variable $z_{n,i}^{u,r}$ is defined as indicating whether  VNF $u \in N_v^r$ of chain $r$ is mapped to instance $i \in I_{t_u }$, of type $t_u$ running on physical node $n \in N_p$ or not. This variable is obtained by the multiplication of $s_{i,t_u}^{u,r}$ and $p_n^{i,t_u }$:
\begin{equation}\label{c6}
z_{n,i}^{u,r} = s_{i,t_u}^{u,r}.p_n^{i,t_u} {\ }{\ }{\ } \forall r \in {G_v},\forall u \in N_r^v,\forall i \in {I_{{t_u}}},\forall n \in {N_p}
\end{equation}
The following constraints linearize \eqref{c6}.
\begin{align}
z_{n,i}^{u,r} \le s_{i,{t_u}}^{u,r}{\ }{\ }{\ } & \forall r \in {G_v},\forall u \in N_r^v,\forall i \in {I_{t_u^r}},\forall n \in {N_p} \label{c7}\\
z_{n,i}^{u,r} \le p_n^{i,{t_u}}{\ }{\ }{\ } & \forall r \in {G_v},\forall u \in N_r^v,\forall i \in {I_{t_u^r}},\forall n \in {N_p} \label{c8}\\
s_{i,{t_u}}^{u,r} + p_n^{i,{t_u}} - 1 \le z_{n,i}^{u,r}{\ }{\ }{\ } & \forall r \in {G_v},\forall u \in N_r^v,\forall i \in {I_{t_u^r}},\forall n \in {N_p} \label{c9}
\end{align}

At the second step, $x_n^ {u,r}$ is defined as another binary variable by taking summation of $z_{n,i}^{u,r}$ on $i$ that indicates whether VNF $u \in N_r^v$ of SFC $r$ is embedded into the physical node $n \in N_p$ or not.
\begin{equation}\label{c10}
\sum\limits_{i \in {I_{{t_u}}}} {z_{n,i}^{u,r} = x_n^{u,r}{\ }{\ }{\ }\forall r \in {G_v},\forall u \in N_r^v,\forall n \in {N_p}}
\end{equation}
For mapping virtual links, we use variable $\delta_{\left(n,m\right)}^{r,\left(u,v\right)}$ which indicates the percentage of the traffic of virtual link ${\left(u,v\right)} \in E_r^v$ of SFC $r$ that is mapped to physical link ${\left(n,m\right)} \in E_p$ of the substrate network.

${bw}_{\left( n,m \right)}$ is a real variable that indicates the amount of the allocated bandwidth on physical link ${\left( n,m \right)} \in E_p$. $\lambda_{ i,t}$ is a real variable which denotes the traffic entering instance $i$ of type $t$. Variable $y_n^{i,t}$ is the entering traffic to the hypervisor queue at node $n \in N_p$ which should be forwarded to  instance $i \in I_t$ of type $t \in T$, so we have
\begin{equation}\label{c11}
y_n^{i,t} = p_n^{i,t}.{\lambda _{i,t}}{\ }{\ }{\ }\forall t \in T,\forall i \in {I_t},\forall n \in {N_p}
\end{equation}
This constraint is linearized as following where $M$ is a big value
\begin{align}
p_n^{i,t}.M \ge y_n^{i,t} {\ }{\ }{\ } & \forall t \in T,\forall i \in {I_t},\forall n \in {N_p} \label{c12}\\
{\lambda_{i,t}} - (1 - p_n^{i,t}).M \le y_n^{i,t} {\ }{\ }{\ }& \forall t \in T,\forall i \in {I_t},\forall n \in {N_p} \label{c13}\\
{\lambda _{i,t}} \ge y_n^{i,t} {\ }{\ }{\ } & \forall t \in T,\forall i \in {I_t},\forall n \in {N_p} \label{c14}
\end{align}

The end-to-end delay of every accepted SFC is the summation of a) the delay in the hypervisors that the SFC enters them which is denoted by $d_r^{hyp}$, b) the delay of the instances that serve the VNFs of the chain which is denoted by $d_r^{ins}$, and c) the propagation delay of the links forwarding the traffic of the chain which is shown by $d^{link}_{r}$.
Variable $l_n^r$ indicates the delay that SFC $r$ faces in passing through physical node $n \in N_p$, it is obtained as
\begin{equation}\label{c15}
l_n^r = a_n^r.{{\bar{d}}_n}{\ }{\ }{\ }\forall r \in {G_v},\forall n \in {N_p}
\end{equation}
where  $\alpha_n^r$ indicates whether or not SFC $r$ passes through physical node $n \in N_p$. This nonlinear constraint is linearized in the same way as $y^{i,t}_n$.

$q_{i,t}^{u,r}$ is another variable that denotes the delay of SFC $r$ since its VNF $u \in N_r^v$ is mapped to instance $i$ of type $t$. It is obtained as:
\begin{equation}\label{c19}
q_{i,t}^{u,r} = s_{i,t}^{u,r}.{{\bar{d}}_{i,t}}{\ }{\ }{\ }\forall r \in {G_v},\forall u \in N_r^v,\forall i \in {I_{t_u^r}},\forall t \in T
\end{equation}
which is linearized as $y^{i,t}_n$.

The $cost$ variable is the total money payed by the service provider for bandwidth allocation on physical links, turning on physical servers, and VNF instances activation. $rev$ is the amount of money that the provider earns from the accepted SFCs, and the $gain$ variable is the difference between $rev$ and $cost$ that indicates the provider’s business profit.

\subsubsection{Objective Function and Constrains}
In this problem, the objective is to maximize the $gain$.
\begin{equation}\label{c23}
  \text{Minimize }gain
\end{equation}
which is computed by the constrains \eqref{c24}-\eqref{c26}:
\begin{align}
gain =& \ rev - cost \label{c24} \\
rev =& \sum\limits_{r \in {G_r}} {re{v_r}.{A_r}} \label{c25}\\
cost =& \sum\limits_{{n} \in {N_p}} {\sigma _n.{\alpha}^r_n}+ \sum\limits_{t \in T} \sum\limits_{i \in {I_t}} {C_t}.x_t^i + {\rho} .\sum\limits_{\left( {m,n} \right) \in {E_p}} {b{w_{\left( {m,n} \right)}}} \label{c26}
\end{align}

As already eluded, a two-level mapping is conducted to deploy SFCs, at the first level  VNFs are mapped to instances and in the second level instances are mapped to physical nodes. In the first level, if a SFC is accepted, all its VNFs must be mapped to an instance, which is formulated as \eqref{c27}.
\begin{equation}\label{c27}
\sum\limits_{i \in {I_{{t_u}}}} {s_{i,t}^{u,r}}  = {A_r}{\ }{\ }{\ }\forall r \in {G_v},\forall u \in N_r^v
\end{equation}
In the second level, an instance must be mapped to a physical node  if at least a VNF is mapped to the instance, which is imposed by constraint \eqref{c28}.
\begin{equation}\label{c28}
s_{i,t}^{u,r} \le \sum\limits_{n \in {N_p}} {p_n^{i,t}}{\ }{\ }{\ }\forall r \in {G_v},\forall u \in N_r^v,\forall i \in {I_{t_u^r}}
\end{equation}

To map virtual links, under the assumption of flow-splitting, a set of paths should be found in the physical network which is obtained by the flow conservation constraint in \eqref{c29}; then the required bandwidth should be allocated, which is formulated by \eqref{c30}, with subject to the capacity of physical links in constraint \eqref{c31}.
\begin{align}
\sum\limits_{\left( {n,m} \right) \in {E_p}} {\delta _{\left( {n,m} \right)}^{r,\left( {u,v} \right)}}  - \sum\limits_{\left( {m,n} \right)} {\delta _{\left( {m,n} \right)}^{r,\left( {u,v} \right)} = x_n^{u,r} - x_n^{v,r}}{\ }{\ }{\ } & \forall r \in {G_v},\forall \left( {u,v} \right) \in E_r^v,\forall n \in {N_p} \label{c29}\\
\sum\limits_{r \in {G_v}} {\sum\limits_{\left( {k,l} \right) \in E_r^v} {w_{\left( {k,l} \right)}^r} } .\delta _{\left( {m,n} \right)}^{r,\left( {u,v} \right)} = b{w_{\left( {n,m} \right)}}{\ }{\ }{\ } & \forall \left( {n,m} \right) \in {E_p} \label{c30} \\
b{w_{\left( {n,m} \right)}} \le {\theta _{\left( {n,m} \right)}}{\ }{\ }{\ } & \forall \left( {n,m} \right) \in {E_p}\label{c31}
\end{align}

The CPU, storage, and memory capacity limitations in physical nodes are respectively imposed by constraints \eqref{c32}, \eqref{c33}, and \eqref{c34}.
\begin{align}
\sum\limits_{t \in T} {\sum\limits_{i \in {I_t}} {CP{U_t}} } .p_n^{i,t} \le \theta _n^{CPU}{\ }{\ }{\ } & \forall n \in {N_p} \label{c32}\\
\sum\limits_{t \in T} {\sum\limits_{i \in {I_t}} {str{g_t}} } .p_n^{i,t} \le \theta _n^{strg}{\ }{\ }{\ }& \forall n \in {N_p} \label{c33}\\
\sum\limits_{t \in T} {\sum\limits_{i \in {I_t}} {me{m_t}} } .p_n^{i,t} \le \theta _n^{mem}{\ }{\ }{\ } & \forall n \in {N_p} \label{c34} 
\end{align}

In order to compute the processing and queuing delays, the input traffic rate for each instance $i \in I_t$ and each physical node $n \in N_p$ are respectively calculated by \eqref{c36} and \eqref{c37}.
\begin{align}
\sum\limits_{r \in {G_v}} {\sum\limits_{u \in N_v^r} {{f_r}} } .s_{i,t}^{u,r} = {\lambda _{i,t}}{\ }{\ }{\ } & \forall t \in T,\forall i \in {I_t} \label{c36}\\
\sum\limits_{t \in T} {\sum\limits_{i \in {I_t}} {y_n^{i,t}} }  = {\lambda _n}{\ }{\ }{\ } & \forall n \in {N_p} \label{c37}
\end{align}
Using the variables $\lambda_{n}$ and $\lambda_{i,t}$ and the given parameters $u_n$ and $u_t$, the queuing and processing delays in the physical nodes and the VNF instances will be as \eqref{c38} and \eqref{c39}, respectively.
\begin{align}
\log \left( {{{\bar d}_n}} \right) + \log \left( {{\mu _n} - {\lambda _n}} \right) \geq 0{\ }{\ }{\ } & \forall n \in {N_p} \label{c38} \\
\log \left( {{{\bar d}_{l,t}}} \right) + \log \left( {{\mu _t} - {\lambda _{i,t}}} \right) \geq 0{\ }{\ }{\ } & \forall t \in T,i \in {I_t} \label{c39}
\end{align}
It should be noted that these constraints are convex constraint.

Constraints \eqref{c43}, \eqref{c44}, and \eqref{c45} calculate $d_r^{hyp}$, $d_r^{ins}$, and $d_r^{link}$ for each SFC $r$ as the imposed average delay from hypervisors, instances, and links respectively. To guarantee the end-to-end delay, the summation of the delays should be less than the given threshold which is imposed by \eqref{c46}.
\begin{align}
\sum\limits_{n \in {N_p}} {l_n^r = d_r^{hyp}}{\ }{\ }{\ } & \forall r \in {G_v} \label{c43} \\
\sum\limits_{u \in N_r^v} {\sum\limits_{i \in {I_{t_u^r}}} {q_{i,t_u^r}^{u,r}} }  = d_r^{ins}{\ }{\ }{\ } & \forall r \in {G_v} \label{c44}\\
\sum\limits_{\left( {n,m} \right) \in {E_p}} {\sum\limits_{\left( {v,u} \right) \in E_v^r} {{d_{\left( {n,m} \right)}}.\delta _{\left( {n,m} \right)}^{r,\left( {u,v} \right)}} }  = d_r^{link}{\ }{\ }{\ } & \forall r \in {G_v} \label{c45}\\
d_r^{hyp} + d_r^{ins} + d_r^{link} \le d_r^{th}{\ }{\ }{\ }\forall r \in {G_v} \label{c46}
\end{align}


Finally, the values of variables $\alpha_n^r$ and $x_n^{u,r}$ need to be determined in order to compute the $cost$, which are obtained by \eqref{c37}-\eqref{c39}.
\begin{align}
x_n^{u,r} \le \alpha_n^r{\ }{\ }{\ } & \forall r \in {G_v},\forall u \in N_v^r,\forall n \in {N_p}\label{c40} \\
\alpha_n^r \le \sum\limits_{u \in N_v^r} {x_n^{u,r}}{\ }{\ }{\ } & \forall r \in {G_v},\forall n \in {N_p} \label{c41}\\
s_{i,t}^{u,r} \le x_t^i{\ }{\ }{\ }& \forall r \in {G_v},\forall u \in N_v^r,\forall i \in {I_t},\forall t \in T \label{c42}
\end{align}

By putting the objective function and the constraints altogether, we have the formulation of the DCSM as following

\begin{center}
\centering
\begin{tabular}{ll}
  \textbf{Model}: & DCSM \\
  \textbf{Objective}: & \eqref{c23}\\
  \textbf{Constraints}: & \eqref{c6}-\eqref{c19}, \eqref{c24}-\eqref{c41}\\
\end{tabular}
\end{center}


This model is a mixed integer convex optimization problem. Unfortunately, it cannot be used as a practical solution of the DCSM problem since the problem is an NP-hard \cite{alameddine2017interplay,bhamare2017optimal,chen2018automated}. In the following section, by decomposing this complex model, we develop an efficient solution for this problem.

\section{Proposed Solution}\label{PS}

This section describes the proposed solution, named MLDG (Multi-Level Delay-Guaranteed) mapping, for the DCSM problem. In the following, we first explain the main ideas and the overall procedure of the solution, then discuss the details of the steps.

Inspecting the problem identified two sources of complexity. The first one is the mapping of the instances on physical nodes, which is formulated by decision variable $p_n^{i,t}$ in the optimization model,  and the second source is the mapping of the virtual functions to the instances, denoted by $s_{i,t}^{u,r}$. Combinations of different values of these variables  yield to the exponential growth of the model solution space, and consequently cause intractability of the model even in small instances.

\begin{figure}
  \centering
  \captionsetup{justification=centering}
  \includegraphics[width=0.45\textwidth]{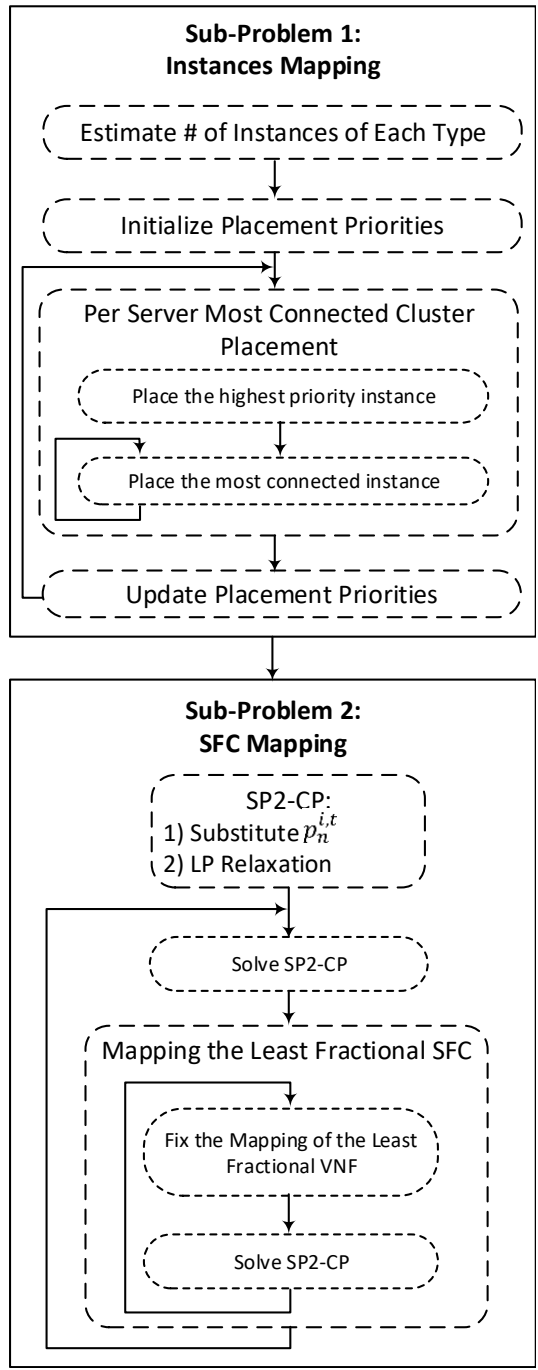}
  \caption{The overall procedure of the proposed solution}\label{Fig4}
\end{figure}

Based on this observation, the main idea to approach the problem is to decompose it into two subproblems. In the first subproblem, the candidate instances of each function type are determined and placed in the physical network. In the second subproblem, instead of considering all SFCs in a large MINLP model, the service chains are deployed one-by-one using an iterative algorithm. This decomposition eliminates the coupling between $p_n^{i,t}$ and $s_{i,t}^{u,r}$ that makes that problem solvable in a polynomial time. The overall procedure of the proposed solution is depicted in \ref{Fig4}; and the details are discussed in the following subsections.

\subsection{Subproblem 1: Instance Mapping}

The main objective of the first subproblem is to determine the value of the decision variables $p_n^{i,t}\ \forall t\in T$,$ \forall i\in I_t$,$\forall n\in N_p$, which is the first source of the complexity of the problem. For this purpose, the instances should be placed in the physical network that needs to answer two questions: 1) what is the required number of instances of each type? and 2) where is the suitable location for the instances in the physical network?

In answering the first question, we should not underestimate the number of instances otherwise it causes rejection of demands in the second subproblem. Therefore, by considering the case where all SFCs are accepted and according to the assumption of instance sharing among VNFs, the estimated number of the instances of the type $t$ is
\begin{equation}\label{c47}
{\eta ^t} = \left\lceil {\frac{{\sum\limits_{\forall r \in {G_v}} {\left( {\sum\limits_{\forall u \in N_v^r|type(u) = t} {{f_r}} } \right)} }}{{\theta .{\mu _t}}}} \right\rceil{\ }{\ }{\ }\forall t \in T
\end{equation}
where $0 < \theta  < 1$ is a parameter to scale down the capacity of the instances. This scaling is needed since in practice, as VNFs cannot be split among multiple instances,  it is not possible to utilize the fully capacity, $ \mu_t$, of the instances. This is the first step of the instance mapping sub-problem in figure \ref{Fig4}.

To answer the second question, we develop an algorithm based on a key observation. As explain, three factors contribute to the end-to-end delay that are the propagation delay of physical links, the operating system and hypervisor delay in physical servers, and the queuing and processing delay in instances. The key point is that if two successive VNFs of a chain are mapped to the instances which are located in the same physical node, then the virtual link connecting these two VNFs resides inside the physical node and consequently the first and second delay factors are eliminated. Moreover, this approach eliminates the cost of bandwidth allocation on physical links. Therefore, in the instance mapping stage, we should seek to find a cluster of instances which are \textit{mostly connected} to each other and place the cluster on a single physical server.

To select the first member of a cluster, we define the placement priorities per type $t$ as following
\begin{equation}
{\pi ^t} = {\tilde \eta ^t}.{\Lambda ^t}{\ }{\ }{\ }\forall t \in T
\end{equation}
where ${\Lambda ^t}$ is the total number of virtual links $\left(u,v \right)$ in all chains that either $u$ or $v$ is of type $t$, and ${\tilde \eta ^t}$ is the number of instances of type $t$ which have not been placed yet. To initialize the priorities, ${\tilde \eta ^t}={\eta ^t}$.

After selecting the first instance of a cluster, subsequent instances are selected according to the number of connections (virtual links) between them and the members of the cluster. More precisely, we define belonginess of type $t$ to cluster $\kappa$ as
\begin{equation}
\beta_{\kappa}^t = \Big| {\left\{ {\left( {u,v} \right),\left( {v,u} \right) \in E_r^v{\ }{\ }\forall r \in {G_v} \text{ s.t. } u \in \kappa \text{ and } type(v) = t} \right\}} \Big|
\end{equation}
At each step, an instance of a type $t$ with the largest $\beta_{\kappa}^{t}$ is selected as a member of cluster if it does not violate the capacity of the physical server. When it is not possible to add a new member to the cluster, this procedure is repeated for a new physical server and its corresponding new cluster. The details of the procedure are explained in algorithm \ref{Algorithm1}.

\begin{small}
\begin{algorithm}
INPUT:${\ }G_p, \eta^t$\\
OUTPUT:${\ }p^{i,t}_{n}$
\begin{algorithmic}[1]
\STATE $\pi ^t=\eta^t\Lambda^t  {\ } \forall t \in T $\\
\STATE $n \leftarrow 0$\\
\WHILE {$ {\ }\exists t \in T {\ } \text{s.t.} {\ }\pi ^ t \ne 0$}
\STATE $\tilde{t} \gets \textbf{argmax}_{t \in T} \{ \pi^t \}$
\STATE $\eta ^ t \leftarrow \eta ^ t - 1$
\STATE $\kappa \gets \{ i {\ } \text{s.t.} {\ }type\left( i \right) = t \}$
\STATE $n \leftarrow n+1$
\STATE $p^{i,t}_{n} \leftarrow 1$
\REPEAT
\STATE $\tilde{t} \gets \textbf{argmax}_{t \in T} \{ \beta ^t_k \}$
\IF {physical server $n$ has enough resources \textbf{and} $\eta ^t \ne 0$}
\STATE $\eta ^ t \leftarrow \eta ^ t - 1$
\STATE $\kappa \gets \{ i {\ } \text{s.t.} {\ } type\left( i \right) = t \} \cup \kappa$
\STATE $p^{i,t}_{n} \leftarrow 1 {\ } $
\ENDIF
\UNTIL{$ {\ }\exists t \in T {\ }\text{s.t.} {\ }\pi ^ t \ne 0$ \textbf{and} server $n$ has sufficient capacity}
\STATE $\pi ^t=\eta^t\Lambda^t  {\ } \forall t \in T $
\ENDWHILE
\end{algorithmic}
\caption{Most Connected Cluster Placement}
\label{Algorithm1}
\end{algorithm}
\end{small}

In this algorithm, the placement priorities are initialized in line 1, the first member of a cluster is found and placed in lines 4-8, similarly in lines 10-14 the most connected instance is added to the cluster, and finally in line 17 the priorities are updated.

\subsection{Subproblem 2: SFC Mapping}

The main steps of the second subproblem are depicted in Figure \ref{Fig4}. In this subproblem, by substituting the values of decision variables of $p_n^{i,t}$, obtained from the first subproblem, in the MIP model and relaxing the binary variables of it, we obtain the convex problem named SP2-CV. While the solution of the convex model is not necessarily a feasible solution of the DCSM, it can be post-processed by rounding techniques to obtain integer solution. Here, we focus on the least fractional VNF mapping variable, $s_{i,t}^{u,r}$, and deploy the chains in two nested loops. The outer loop, in each iteration, finds the chain with the minimum fractional VNF mapping, and the inner loop processes the VNFs of the chain. In each iteration of the loop, the minimum fractional $s_{i,t}^{u,r}$ is selected, and rounded to 1. After fixing the variable, the SP2-CV model is solved to update the remaining variables. This procedure continues until either all VNFs of the chain is mapped, or the chain is rejected due to infeasibility of the model. The subsequent chains are deployed in the same way. The details of these steps are explained in algorithm \ref{Algorithm2}.

\begin{small}
\begin{algorithm}
INPUT:$G_p, G_v, p^{i,t}_{n}, T$\\
OUTPUT:$A_{r}, s^{u,r}_{i,t}, p^{i,t}_{n}$
\begin{algorithmic}[1]
\REPEAT
\STATE $s^{u,r}_{i,t} \gets \text{solve }SP2-CV$
\STATE ${\theta}_{r} \gets \sum\limits_{u \in {N_r^v}} \textbf{max}_{i \in I_{t_u}} \left( \{ s^{u,r}_{i,t} \} / |N^v_r| \right)$
\STATE $\tilde{r} \leftarrow \textbf{argmax}_{u \in N^v_{\tilde{r}}} \{ \theta_r \}$
\STATE $A_{\tilde{r}} \leftarrow 1$

\WHILE{$ {\ }\exists u \in N^{v}_{\tilde{r}} {\ } \textbf{s.t.} {\ }s^{u,\tilde{r}}_{i,t} \ne 1$}
\STATE $\tilde{u} \gets \textbf{argmax}_{ u \in N^{v}_{\tilde{r}}} \{ s^{u,\tilde{r}}_{i,t} \}$
\STATE $s^{\tilde{u},\tilde{r}}_{i,t} \leftarrow 1$
\STATE $\text{Substitute} {\ } s^{\tilde{u},\tilde{r}}_{i,t} {\ } \text{in} {\ } SP2-CV$
\STATE $\bar{s}^{u,r}_{i,t} \leftarrow \text{solve} {\ } SP2-CV$
\IF {$\bar{s}^{u,r}_{i,t} {\ }\text{is infeasible}$}
\STATE $s^{u,\tilde{r}}_{i,t} \leftarrow 0 {\ } \forall u \in N^v_{\tilde{r}}$
\STATE $A_{\tilde{r}} \leftarrow 0$
\STATE $\textbf{break}$
\ELSE
\STATE $s^{u,r}_{i,t} \leftarrow \bar{s}^{u,r}_{i,t} $
\ENDIF
\ENDWHILE
\UNTIL {$\exists r \in G_v$}
\end{algorithmic}
\caption{SFC Mapping}
\label{Algorithm2}
\end{algorithm}
\end{small}

In line 2, the SP2-CV model, wherein the $s_{i,t}^{u,r}$ variables are not necessarily binary, is solved. In line 3, the least fractional $s_{i,t}^{u,r}$ is found regarding to length of its corresponding SFC, and in line 4 the mentioned SFC, which is denoted by $\tilde r$, is identified.  It is assumed that this chain will be accepted at line 5, and in the inner loop, the VNFs of the chain are investigated. At each iteration, the $u\in N_{\tilde r}^v$ with least fractional value $s_{i,t}^{u,\tilde r}$ is identified in line 7. It is rounded to 1 in line 8. After substituting this value in the model, it is solved in line 10. If it is an infeasible model, this SFC is rejected in lines 11-14, otherwise, $s_{i,t}^{u,r}$ values are update and the loops continues to map the remaining VNFs of SFC $\tilde r$. The outer loop continues until all SFCs are either accepted or rejected.

\subsection{Complexity Analysis}

The computational complexity of the proposed solution is analyzed in this section. In the first subproblem, the complexity of computing $\eta ^t$ is $O \left( \left|T \right| \left| G_v\right|N^* \right)$ where $N^*$=max $\left| N_v^r \right|$. Initializing $\pi^t$ takes $O \left( T \right)$ time. Placement of instances on physical server, the two nested loops in lines 3-17 of algorithm \ref{Algorithm1}, is performed in $O \left( H \right)$  iterations where $H=\sum \eta ^t $. Operations in lines 3-17 are $O \left( 1 \right) $ except the argmax operation and updating $\pi ^t$ which take $O \left( T \right) $. Therefor, the complexity of the first subproblem is
\begin{equation*}
 O(\text{subprobelm-1}) = O( | T | | G_v | N^* ) + O( T ) + O( HT) = O(| T| | G_v| N^* )+ O( HT )
\end{equation*}

In the second subproblem, the nested loops and solving the SP2-CV model are the main contributors of the complexity. The complexity of the outer loop is
\begin{equation*}
O(\text{subproblem-2}) = O( | G_v | ( CV ) + R^* + N^* O ( \text{inner} ) )
\end{equation*}
Where $O(CV) $ and $O(\text{inner})$ are respectively the complexity of convex programming and the complexity of lines 6-15 of algorithm \ref{Algorithm2}. There are efficient algorithms for convex programming \cite{nemirovski2004interior}. The complexity of lines 6-15 is
\begin{equation*}
O(\text{inner})= O (N^*+ O(LP)+N^*)
\end{equation*}
So, we have
\begin{equation*}
O(\text{subproblem-2}) = O ( |G_r |  N^*  O( CV ))
\end{equation*}
By putting these analyses together, the complexity of the proposed method is
\begin{equation*}
O(| G_r | N^* O (CV)) +O(|T| |G_v | N^*)+ ( HT ) 
\end{equation*}
which is polynomial in terms of the size of substrate network, the number of function types, the number of SFCs and their sizes.

\section{Simulation Results}\label{SR}
In this section, at first, the  simulation settings are discussed in detail, then comparisons between the proposed solution and the optimal solution, and the algorithm proposed in \cite{fischer2019construction} are presented; and finally, the effects of the parameters of the proposed algorithm are analyzed.

\subsection {Simulation Settings}\label{SS}
The system used for the simulations has a Core $i7$ Intel processor with 2.7 GHz frequency, 16 GB of RAM, and 512 GB of SSD hard disk. The simulation environment  is implemented by Python, and ANACONDA Package \cite{anaconda} is used to configure the SCIP solver \cite{GamrathEtal2020OO}. Since the problem studied in \cite{fischer2019construction}  is the most similar one to our problem, we implemented its proposed algorithm, named Annealing Based Simulation Approach (ABSA), and compared its performance to the proposed solution. It should be noted that the problems are not exactly the same, so, we modified them in such a way to make a fair comparison between two solutions. The changes are discussed in the next sections.

\begin{figure}
  \centering
  \captionsetup{justification=centering}
  \includegraphics[width=0.4\textwidth]{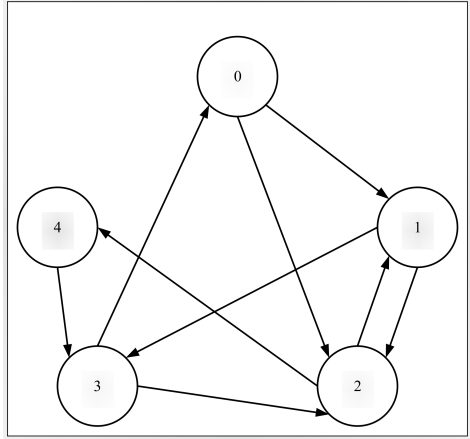}
  \caption{The Small Topology Used for Comparison with The Optimal Solution.}\label{Fig5}
\end{figure}

In this paper, two topologies are used as the substrate network. Figure \ref{Fig5} shows the small topology, used for the of proposed solution against to the optimal solution. In this topology, the bandwidth of all the links equal 1000 MB. The CPU capacity of all the nodes are randomly chosen in [220, 260] cores. Storage capacity of all nodes is equal to 4000 GB and memory capacity is 1000 GB. Table \ref{tt8} and Table \ref{tt9} show the details of input SFCs and VNF types which are used in the topology shown in Figure \ref{Fig5}.

\begin{table} 
\begin{center}
\caption{The Setting of the SFCs Used for Comparison with the Optimal Solution in Topology of Fig. \ref{Fig5}.}
\label{tt8}
\centering
\small\addtolength{\tabcolsep}{0pt}
\scalebox{0.9}{
\begin{tabular}{|l|c|}
  \hline
  Parameter & Value \\
  \hline  
  \hline
   VNF types $\left( t^r_u \right)$ & Randomly in \cite{hawilo2014nfv,han2015network,herrera2016resource,yu2017qos}\\
  \hline
   VNF Number per SFC $(|N^v_r|)$ & 3 VNFs\\
  \hline
     Bandwidth $(f_r)$ & 30 MB\\
  \hline
    Revenue $ \left( {rev}_r \right)$ & 9000\\
  \hline
      End-to-end delay threshold $ \left( d^{th}_r \right)$ & 1 $\mu$s \\
  \hline
\end{tabular}
}
\end{center}
\end{table}

\begin{table} 
\begin{center}
\caption{The Settings of the VNF Types Used for Comparison with the Optimal Solution in Topology of Fig. \ref{Fig5}.}
\label{tt9}
\centering
\small\addtolength{\tabcolsep}{0pt}
\scalebox{0.9}{
\begin{tabular}{|c|c|c|c|c|}
  \hline
  Parameter & \multicolumn{4}{c|}{Value per Type}\\
  \hline 
  \hline
$t$ & 1 & 2 & 3 & 4\\
\hline
${cpu}_t$ (cores) & 80 & 50 & 50 & 60\\
\hline
${strg}_t$ (GB) &200&250&250&300\\
\hline
${mem}_t $ (GB) &160&100&100&120\\
\hline
${\mu}_t $ (MBps) &100&100&100&100\\
\hline
$C_t$ &500&600&700&500 \\
\hline
\end{tabular}
}
\end{center}
\end{table}

\begin{table} 
\begin{center}
\caption{NFVI Setting in BtEurope Topology Used for Comparison with ABSA.}
\label{tt10}
\centering
\small\addtolength{\tabcolsep}{0pt}
\scalebox{0.9}{
\begin{tabular}{|c|c|}
  \hline
  Parameter & Value \\
  \hline
  \hline
   $ \theta^{CPU}_n$ & 24 cores\\
  \hline
   $ \theta^{strg}_n $ & 10000 GB\\
  \hline
    $ \theta^{mem}_n $  & 256 GB\\
  \hline
   $\theta_{ \left( n,m \right)}$  & 10000 MB\\
  \hline
     $\rho$  & 10\\
  \hline
     $\sigma_{n}$  & 1000\\
  \hline
  $ \mu_n$ & 2000 {MB}_{ps} \\
  \hline
    $ d_{\left( n,m \right)}$ & 0.5 $\mu$s\\
  \hline
    $d_n$ & 3 $\mu$s \\
  \hline
\end{tabular}
}
\end{center}
\end{table}

Comparisons to the proposed algorithm in \cite{fischer2019construction} were conducted in the BtEurope topology, from The Internet Topology Zoo, available at \cite{6027859}. The settings are  summarized in Table \ref{tt10}. The input SFCs and the VNF types used for these comparisons are respectively shown in Table \ref{tt11} and Table \ref{tt12}.

\begin{table} 
\begin{center}
\caption{The Setting of the SFCs Used for Comparison with ABSA in BtEurope.}
\label{tt11}
\centering
\small\addtolength{\tabcolsep}{0pt}
\scalebox{0.9}{
\begin{tabular}{|l|c|}
  \hline
  Parameter & Value \\
  \hline
  \hline
   VNF Types $\left( t^{r}_u \right)$ & Randomly in \cite{hawilo2014nfv,han2015network,herrera2016resource,yu2017qos} \\
  \hline
   VNF Number per SFC $\left( | N^v_r | \right)$ & 4 VNFs\\
  \hline
    Bandwidth $(f_r)$  & 100 MB\\
  \hline
  Revenue $\left( {rev}_r \right)$  & $3000|N^v_r|+ 15 f_r (|N^v_r|-1)$\\
  \hline
     End-to-end delay threshold $ \left( d^{th}_{r} \right)$  & 3${\mu}s$\\
  \hline
\end{tabular}
}
\end{center}
\end{table}

\begin{table} 
\begin{center}
\caption{The Settings of the VNF Types Used for Comparison with ABSA in BtEurope.}
\label{tt12}
\centering
\small\addtolength{\tabcolsep}{0pt}
\scalebox{0.9}{
\begin{tabular}{|c|c|c|c|c|}
  \hline
  Parameter & \multicolumn{4}{c|}{Value per Type}\\
  \hline 
  \hline
$t$ & 1 & 2 & 3 & 4\\
\hline
${CPU}_t $ (cores) &4&7&6&5\\
\hline
${strg}_t $ (GB) &90&90&120&100\\
\hline
$ {mem}_t $ (GB) &16&24&32&24\\
\hline
${\mu}_t $ (MBps) &450&500&400&450\\
\hline
$C_t$ &2000&2200&1800&2500\\
\hline
\end{tabular}
}
\end{center}
\end{table}

In these simulations, in addition to the $gain$ which is the main objective of DCSM, we also evaluate the results in terms of acceptance rate. Since the problem is NP-Hard, it is extremely time consuming to reach the exact optimal solution using a solver; hence,  we set the optimality gap threshold equals to 5\%.

\subsection{Comparison with Optimal Solution}
\begin{figure}
  \centering
  \captionsetup{justification=centering}
  \includegraphics[width=0.6\textwidth]{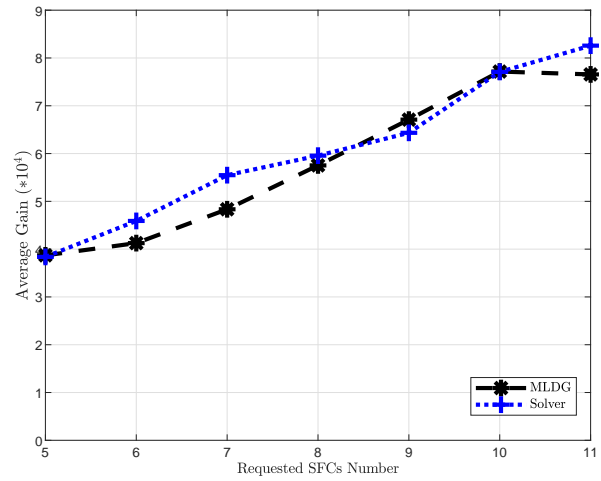}
  \caption{The Average Gain of the Solver and MLDG with Respect to the Number of SFCs.}\label{Fig6}
\end{figure}

\begin{figure}
  \centering
  \captionsetup{justification=centering}
  \includegraphics[width=0.6\textwidth]{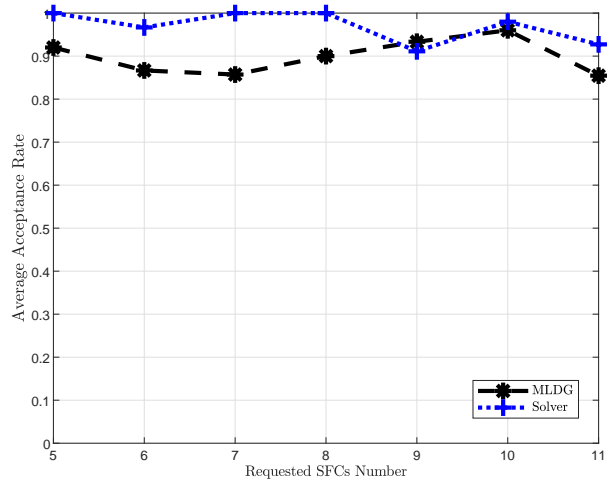}
  \caption{The Average Acceptance Rate of the Solver and MLDG with Respect to the Number of SFCs.}\label{Fig7}
\end{figure}

\begin{figure}
  \centering
  \captionsetup{justification=centering}
  \includegraphics[width=0.6\textwidth]{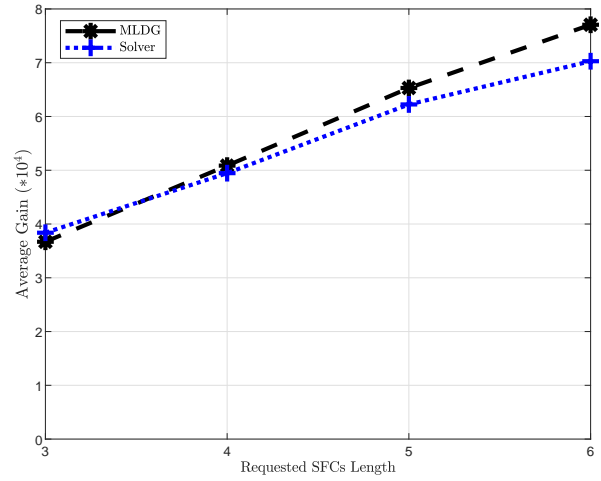}
  \caption{The Average Gain of the Solver and MLDG with Respect to the SFC Length.}\label{Fig8}
\end{figure}

\begin{figure}
  \centering
  \captionsetup{justification=centering}
  \includegraphics[width=0.6\textwidth]{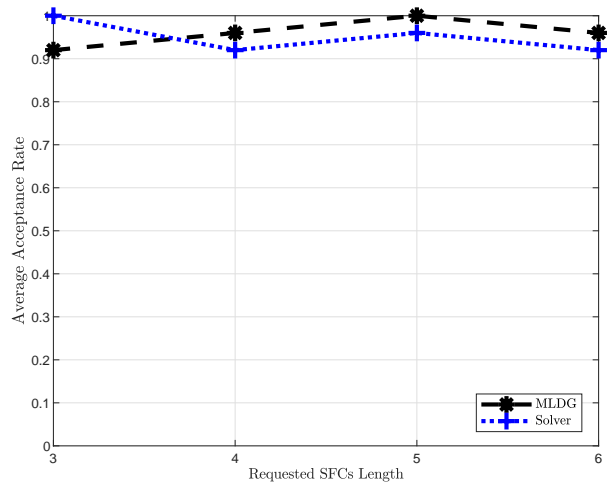}
  \caption{The Average Acceptance Rate  of the Solver and MLDG with Respect to the SFC Length.}\label{Fig9}
\end{figure}

In this section, the efficiency of MLDG is evaluated by comparing it against the solver’s 5\% sub-optimal solution with respect to the number of requested SFCs and the length of SFCs. These simulations were conducted in Figure \ref{Fig5} topology using the configurations presented in the previous subsection. The results are shown in Figures \ref{Fig6}$-$\ref{Fig9}. Figures \ref{Fig6} and \ref{Fig7} show the evaluation results with respect to the number of SFCs; the effect of the length of SFC is evaluated in Figures \ref{Fig8} and \ref{Fig9}.

As it seen the gain increases by increasing the number of requested SFCs and the length of SFC as in this case the algorithms have more opportunity to accept demands and each accepted demand makes more profit. However, the acceptance rate is not necessarily improved by increase the parameters because the algorithms may decide to reject some demands in favor of maximizing the gain. 

These results show that the differences between MLDG and the solver solution are negligible and in some cases, the MLDG solutions are even better than the solver’s 5\% sub-optimal ones. That implies MLDG provides near-optimal solutions. To evaluate this algorithm in a more realistic topology setting, we compare it against another practical algorithm proposed in \cite{fischer2019construction}.



\subsection{Comparison to ABSA}
To the best of our knowledge, the problem investigated in \cite{fischer2019construction} is the most similar one to this paper. Nevertheless, these two problems have some differences which should be manipulated in such a way that makes a fair comparison. Their most considerable differences and the way we handled them are as follows:
\begin{enumerate}
  \item The objective function: the objective of paper \cite{fischer2019construction} is cost minimization, while the objective of DCSM is gain maximization. Based on equation \eqref{c24}, the gain is calculated by subtracting the cost from the revenue. Hence, we defined a revenue for each SFC and changed the objective of  \cite{fischer2019construction} from cost minimization to gain maximization.
  \item Cost calculation: paper \cite{fischer2019construction} does not consider the cost of instance activation for each VNF, therefore, we changed the cost function to consider the cost of activating an instance.
  \item End-to-end delay calculation: In \cite{fischer2019construction} authors have considered a constant delay for the nodes and links. However, in this paper, as illustrated in inequality \eqref{c46}, queuing delay is also considered. We changed our delay model and simplified it as \cite{fischer2019construction}.
\end{enumerate}



As mentioned, ABSA is based on the Simulated Annealing approach, in these simulations, the corresponding parameters are follows: temperature is 1000, cooling rate is 0.05 and the $\lambda$ parameter is 3.

\begin{figure}
  \centering
  \captionsetup{justification=centering}
  \includegraphics[width=0.6\textwidth]{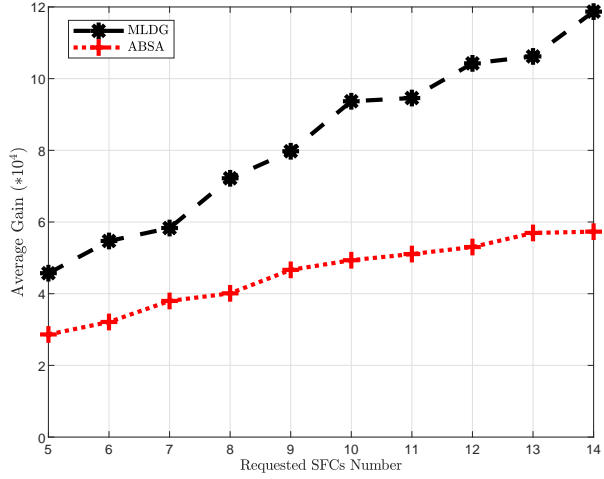}
  \caption{The Average Gain of MLDG and ABSA with Respect to the Number of SFCs.}\label{Fig10}
\end{figure}

\begin{figure}
  \centering
  \captionsetup{justification=centering}
  \includegraphics[width=0.6\textwidth]{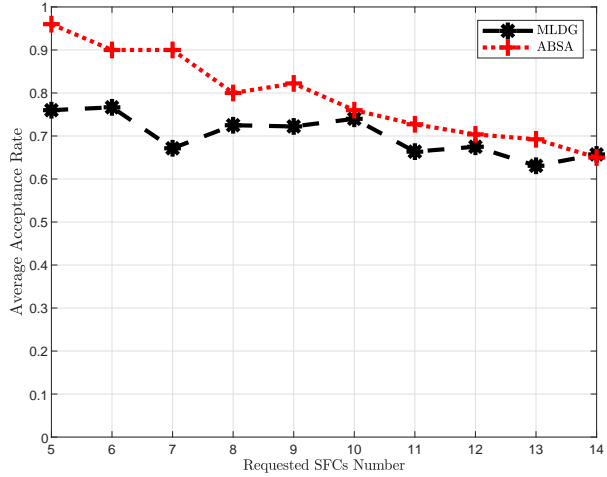}
  \caption{The Average Acceptance Rate of MLDG and ABSA with Respect to the Number of SFCs.}\label{Fig11}
\end{figure}

Figure \ref{Fig10} and Figure \ref{Fig11} show the average gain and the average acceptance of the proposed solution and ABSA with respect to the number of requested SFCs. Although the acceptance rate of the MLDG algorithm is a bit lower than ABSA, it skillfully accepts a subset demands that make more gain that is justified by the considerable gap between MLDG and ABSA in Figure \ref{Fig10}.

By increasing the number of input SFCs the average gain of both algorithms increase but with different rates. By the increase of the number of input SFCs, the required instances capacity increases which leads to more instances sharing possibility. In the proposed solution, by exploiting this possibility, the instance cost breaks down among multiple chains, and consequently, the average cost of each SFC decreases, and as a result, the average gain increases. On the other hand, ABSA does not share the instances that causes a high cost and the lower gain.

\begin{figure}
  \centering
  \captionsetup{justification=centering}
  \includegraphics[width=0.6\textwidth]{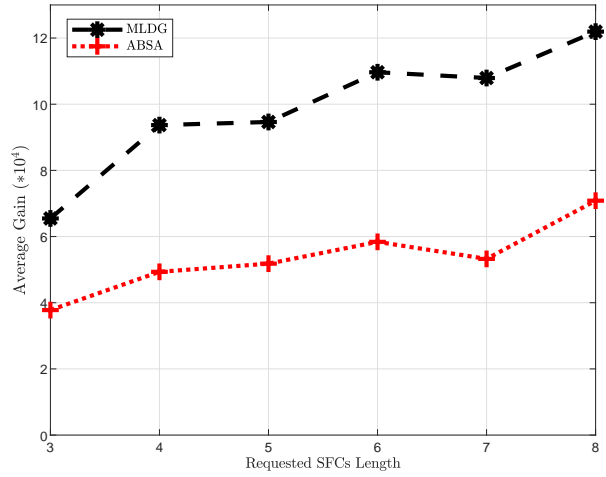}
  \caption{The Average Gain of MLDG and ABSA with Respect to the SFCs Length.}\label{Fig12}
\end{figure}

\begin{figure}
  \centering
  \captionsetup{justification=centering}
  \includegraphics[width=0.6\textwidth]{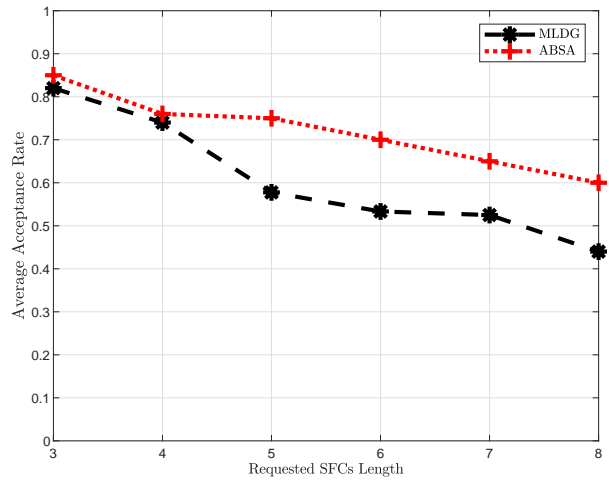}
  \caption{The Average Acceptance Rate of MLDG and ABSA with Respect to the SFCs Length.}\label{Fig13}
\end{figure}

Figure \ref{Fig12} and Figure \ref{Fig13} depict the effect of the length of SFCs on the performance of the algorithms. In these simulations, 10 SFCs are requested. As it shown, SFCs length and gain are positively correlated. Since by increasing the number of VNFs of each SFC, the profit of the accepted demands increases. Moreover, in this case, more instances can be shared among chains that leads to cost breakdown and gain increment. On the other hand, as depicted in figure \ref{Fig13}, the acceptance rate and the SFCs length are negatively correlated. The capacity of each physical node is limited, therefore, longer SFCs should be split in several physical nodes. In this case, several physical nodes should be connected to each other that incurs additional delay that goes beyond the required end-to-end-delay threshold; and consequently the request is rejected.


Putting the results together, indicates that the proposed approach both theoretically and practically is an efficient  solution of the DCSM problem.

\subsection{Design Parameter Impact Analysis}
\begin{figure}
  \centering
  \captionsetup{justification=centering}
  \includegraphics[width=0.6\textwidth]{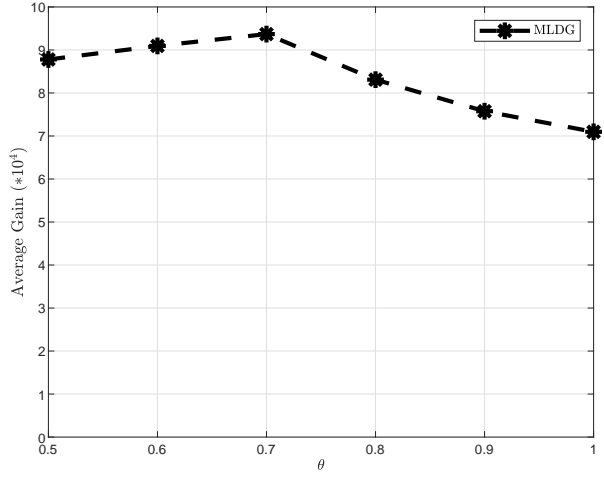}
  \caption{The Average Gain with respect to $\theta$}\label{Fig14}
\end{figure}

\begin{figure}
  \centering
  \captionsetup{justification=centering}
  \includegraphics[width=0.6\textwidth]{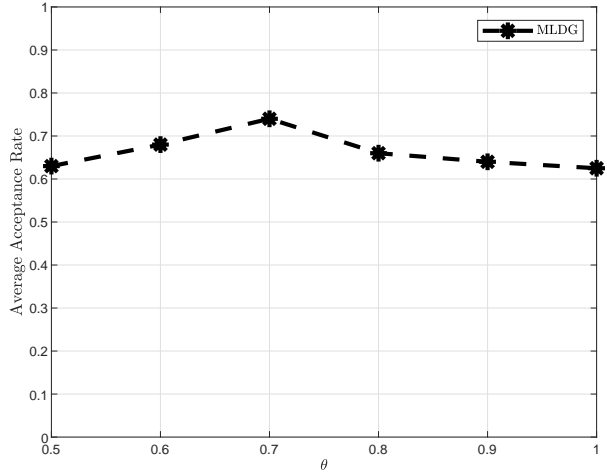}
  \caption{The Average Acceptance Rate with Respect to $\theta$}\label{Fig15}
\end{figure}

In this subsection, the impact of the parameter $\theta$ on the performance of the algorithm is evaluated. The results are depicted in Figure \ref{Fig14} and Figure \ref{Fig15}. According to these figures, there is an optimum value of this parameter, which is 0.7 in our settings. This parameter has a twofold effect on the acceptance rate and gain. When the value of the parameter is much less than the optimum value, according to the \eqref{c47},  the number of instances is  overestimated that leads to high cost and consequently decrease in gain. However, when the value of the parameter is much greater than the optimum value, it causes over utilization of the instances that consequently increase the queuing delay and as a result some demands are rejected due to the end-to-end delay constraint. This decreases the acceptance rate as well as the gain.

\section{Conclusion}\label{CON}
In this paper, the DCSM problem that aims to maximize the profit gain of the service provider while satisfying the QoS requirement including the end-to-end delay is investigated. We presented an accurate modeling of the end-to-end delay that considers the queuing delay as well as propagation delay. The problem is formulated as a mixed integer convex problem, which is used to obtain a performance bound. To approach the problem, a two-level mapping algorithm is proposed by decomposing the optimization problem. At the first level, it maps functions to VNF instances by exploiting the instance sharing potential. In the second level, the instances are deployed in the physical network. The simulation results showed the solution is near optimal.

\bibliographystyle{elsarticle-num}
\bibliography{ref}

\end{document}